# Differentially-Enhanced Sideband Imaging via Radio-frequency Encoding

A. M. Fard,[1, a), b)] A. Mahjoubfar,[2,3] and B. Jalali [2, 3, 4, 5, c)]

[1] Harvard Medical School, Wellman Center for Photomedicine, Massachusetts General Hospital, Boston, MA 02114, USA

[2] Department of Electrical Engineering, University of California Los Angeles, CA 90095, USA

[3] California NanoSystems Institute, Los Angeles, CA 90095, USA

[4] David Geffen School of Medicine, University of California Los Angeles, CA 90095, USA

[5] Department of Bioengineering, University of California Los Angeles, CA 90095, USA

We present a microscope paradigm that performs differential interference imaging with high sensitivity via optical amplification and radio-frequency (RF) heterodyne detection. This method, termed differentially-enhanced sideband imaging via radio-frequency encoding (DESIRE), uniquely exploits frequency-to-space mapping technique to encode the image of an object onto the RF sidebands of an illumination beam. As a proof-of-concept, we show validation experiment by implementing radio frequency ($f$ = 15 GHz) phase modulation in conjunction with spectrally-encoded laser scanning technique to acquire one-dimensional image of a barcode-like object using a commercial RF spectrum analyzer.

a) Electronic mail: mfard.ali@mgh.harvard.edu
b) This research was performed while A. M. Fard was at Department of Electrical Engineering, University of California Los Angeles, CA 90095, USA
c) Electronic mail: jalali@ucla.edu

Optical microscopes are indispensable diagnostic tools for a variety of applications: from industrial production lines[1-3] to advanced clinical and medical settings.[4-6] In particular, imaging modalities capable of visualizing fine surface features and defects of a low-contrast (i.e., semi-transparent) object are highly desirable. The image contrast is often enhanced by either manipulating the object under test, or by exploiting novel illumination and/or detection techniques. For instance, in conventional imaging modalities employing charge-coupled device (CCD) and complementary metal-oxide-semiconductor (CMOS) sensor arrays, image contrasts are often improved by use of contrast agents[7] or other perturbative labeling techniques.[8] Nonetheless, the challenge is fundamentally in collecting sufficient number of photons required to overcome inherent noise (thermal and/or read-out) of the image sensor.[9,10] This technological limitation indeed can be partially overcome by optical image amplification. It is however not readily available in settings employing conventional sensors. Other methods to enhance the image contrast and to highlight otherwise invisible features (e.g., lines and edges) are techniques such as phase-contrast microscopy,[11,12] differential interference contrast microscopy (DICM),[13,14] and oblique illumination microscopy.[15]

In this paper, we report on an imaging modality that performs scanning with high sensitivity and high contrast. Our technique, termed differentially-enhanced sideband imaging via radio-frequency encoding (DESIRE) maps an image to the radio-frequency (RF) sidebands of an illumination beam using spatial dispersion and captures the image using a single-pixel photo-detector and an RF heterodyne receiver. This method also enables use of optical amplification in conjunction with double heterodyne detection of the image – providing superior sensitivity to surface variations of the object compared with conventional imaging techniques. By modulating the illumination beam with an RF signal, a simple and readily-adaptable differential interference microscope can be also realized. As a proof-of-concept demonstration, we also show one-dimensional image of a barcode-like object using a 15-GHz-modulated illumination beam and a commercial RF spectrum analyzer.

DESIRE camera uses a continuous-wave optical source (referred to as "carrier") followed by a phase and/or intensity electro-optic (E/O) modulator to generate the illumination beam as depicted in Figure 1(a). The phase/intensity modulator is fed by an RF modulating signal, creating modulation sidebands in the illumination beam. By performing frequency-to-space mapping using spatial light dispersers such as diffraction grating, the optical carrier and its sidebands are mapped into adjacent points in the target plane. The spatially-dispersed beam is then focused on the object under test such that adjacent points are illuminated by a certain wavelength and its modulation sidebands. After traveling through the object or reflecting back from the object, the spatially-encoded beam is then spatially compressed using the spatial light disperser. The image-encoded sideband-modulated beam is directed to a single-pixel photo-detector. At the photo-detector, the sidebands and the carrier beat with each other, resulting in a signal with amplitude associated with the relative difference (or imbalance) between the RF sidebands. Hence, the resultant signal is the differential interference obtained from two adjacent points on the object. Note that the proposed technique fundamentally differs from our previous demonstration of a high-speed differential interference contrast microscopy,[16] which uses a broadband optical source in conjunction with spectrally-encoded imaging and dispersive Fourier transformation.[17-22] In turn, the proposed imaging technique performs differential interference imaging, while places the burden on the tunable RF components.

A key feature of our technique is the use of double heterodyne detection - offering very high sensitivity in image acquisition. The term "double heterodyne" refers to optical detection in a photo-detector followed by electronic heterodyne receiver using an electronic mixer in combination with a local RF oscillator. In addition, optical amplification can be employed to improve the sensitivity, if the image detection is not shot-noise limited. Note that our proposed method that is essentially a confocal phase microscope fundamentally differs from a recently-developed radio-frequency-tagged fluorescence microscope[23] in which optical heterodyne detection is used to acquire images.

In order to perform non-mechanical scanning in one or two dimensions, the frequency of the RF modulation signal ($f_s$) varies; moving the sidebands across the object, while the wavelength of the illumination beam is constant. As shown in Figure 1(b), the imbalance between the sidebands caused by the object generates a beating RF signal at the photo-detector. Alternatively, as shown in Figure 1(c), non-mechanical scanning may be also performed by sweeping the frequency of the illumination beam ($f_0$), while the RF signal is constant. In latter implementation, the frequency of the modulation signal primarily determines the lateral distance between the two adjacent points for differential interference microscopy.

To validate our DESIRE method, we constructed an experimental apparatus similar to Figure 1(a). The optical source was a CW laser with a center wavelength tunable between 1580 nm to 1600 nm. The CW light was modulated by a 15-GHz RF signal via an E/O phase modulator (EOSpace, X-cut, 20Gb/s). A polarization controller before the modulator was used to maximize the modulation efficiency. The modulated light was then directed to a pair of diffraction gratings (1200 grooves/mm) for spatial-dispersion and collimation into 1D scanning beam. Prior to illumination onto the object, the beam was re-sized using a pair of telescopic lens. An objective lens (Mitutoyo Plan Apo NIR, NA = 0.4) then focuses the illumination beam onto the target plane. As a result, the field of view is found to be ~ 4 mm. In order to avoid saturation of the photo-detector, carrier suppression technique was implemented by biasing the E/O modulator at its minimum transmission. This indeed maintains the same high power level for the RF sidebands, while reducing the power of the carrier. A barcode-like sample (011001010) mounted on a mirror was used as the object for the validation experiments. The reflected light from the sample was directed via an optical circulator toward a high-speed photo-detector (Discovery Semiconductor, R404-HG) followed by an RF heterodyne receiver. The RF heterodyne receiver was a commercial 40-GHz RF spectrum analyzer (HP-8564E). To acquire the image, we chose to capture the beating tone at 30 GHz, which is equal to the frequency distance between the two RF sidebands. This allows us to image two adjacent points that are located at the sidebands' positions and to suppress any information contained by the carrier. The wavelength of the illumination was detuned (1580 nm to 1600 nm) to scan across the sample. Figure 2 shows the reconstructed image (red trace) of the barcode sample. The vertical axis corresponds to the amplitude of the RF interference signal acquired by the RF spectrum analyzer.

We note that while the diffraction grating disperses the light in target plane, it also introduces temporal dispersion, which results in a spurious (unwanted) amplitude modulation through phase modulation to amplitude modulation conversion. This is due to the fact that different wavelengths of the light travel different distances when dispersed in space. Since this spurious amplitude modulation is constant and independent of the object's pattern, it will appear as a background, which is removed by digital signal processing. Alternatively, dispersion compensating modules can be used to compensate for temporal dispersion and to minimize the spurious amplitude modulation.

In summary, we presented an imaging modality, termed differentially-enhanced sideband imaging via radio-frequency encoding (DESIRE) that performs differential interference imaging via double heterodyne detection and frequency-to-space mapping. As a proof-of-concept demonstration, we have shown validation experiments by implementing radio frequency (15-GHz) phase modulation of the illumination beam in conjunction with non-mechanical laser scanning technique to acquire one-dimensional image of a barcode-like object.

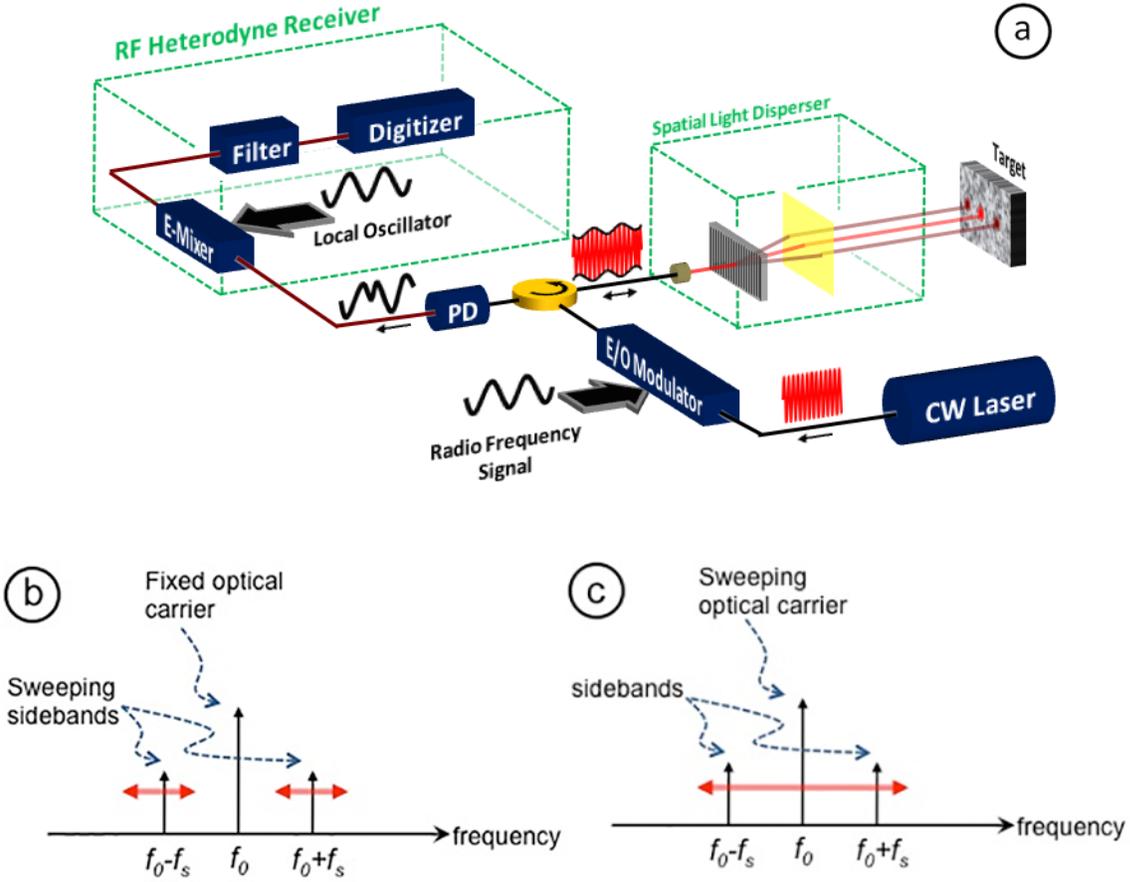

FIG. 1. (a) Conceptual diagram of the differentially-enhanced sideband imaging via radio-frequency encoding (DESIRE). In this imaging method, the illumination beam is modulated by an RF signal. The image is then encoded onto the sidebands of the illumination beam by using wavelength-to-space mapping. Use of optical heterodyne in a photo-detector followed by radio frequency heterodyne receiver enables image acquisition with very high sensitivity. (b) Non-mechanical scanning is performed by sweeping the RF sidebands ($f_s$). (c) Non-mechanical scanning is performed by detuning the wavelength of the illumination beam ($f_0$). CW: Continuous wave, E/O: Electro-optic, PD: Photo-detector.

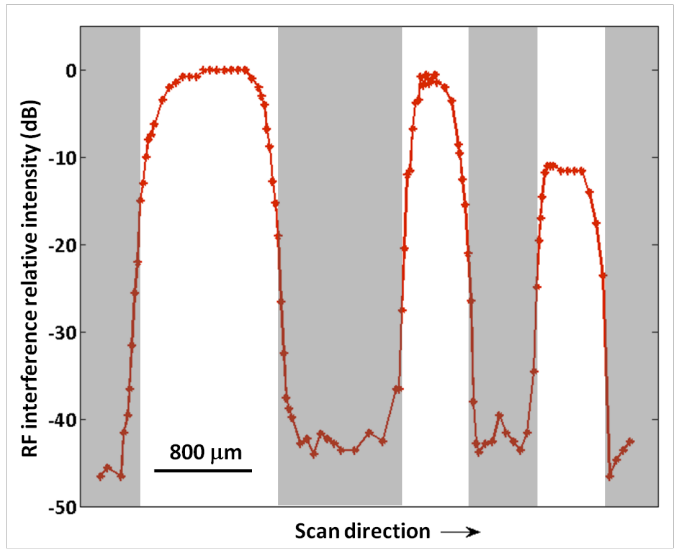

FIG. 2. Red trace shows one-dimensional image of a 011001010 barcode-like sample (grayscale background) mounted on a mirror. To acquire the image, the wavelength of the illumination beam is detuned to scan across the sample.